%% file: output.tex
\documentclass{article}

\usepackage{arxiv}
\usepackage{multirow}
\usepackage{hyperref}
\usepackage[utf8]{inputenc} 
\usepackage[T1]{fontenc}    
\usepackage{hyperref}       
\usepackage{url}            
\usepackage{booktabs}       
\usepackage{amsfonts}       
\usepackage{nicefrac}       
\usepackage{microtype}      
\usepackage{lipsum}
\usepackage{graphicx}
\graphicspath{ {./images/} }

\usepackage{subcaption}
\usepackage{mathtools}
\usepackage{floatrow}

\title{Interpretable Self-supervised Multi-task Learning \\for COVID-19 Information Retrieval and Extraction}

\date{} 					

\author{ \href{https://orcid.org/0000-0003-3479-2774}{\includegraphics[scale=0.06]{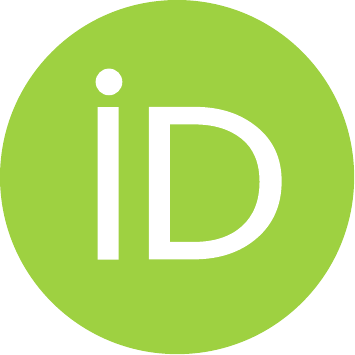}\hspace{1mm}Nima ~Ebadi} \\
	Department of Electrical and Computer Engineering \\
	University of Texas at San Antonio \\
	San Antonio, TX 78249 \\
	\texttt{nima.ebadi@utsa.edu} \\
	\And
	\href{https://orcid.org/0000-0001-9671-577X}{\includegraphics[scale=0.06]{orcid.pdf}\hspace{1mm}Peyman ~Najafirad} \\
	Department of Information Systems and Security\\
	University of Texas at San Antonio \\
	San Antonio, TX 78249 \\
	\texttt{peyman.najafirad@utsa.edu} \\
}

\begin{document}
\maketitle



\maketitle

\input{0_abstract.tex}

\input{1_introduction.tex}
\input{3_methodology.tex}
\input{4_experiments.tex}

\input{5_conclusion.tex}


\section*{FUNDING}
The authors gratefully acknowledge the use of the services of Jetstream cloud, funded by National Science Foundation (NSF) awards 1445604, and the Cloud Technology Endowed Professorship.

\bibliographystyle{unsrt}
\bibliography{anthology,acl2021}







\end{document}

%% file: 0_abstract.tex
\begin{abstract}
The rapidly evolving literature of COVID-19 related articles makes it challenging for NLP models to be effectively trained for information retrieval and extraction with the corresponding labeled data that follows the current distribution of the pandemic. On the other hand, due to the uncertainty of the situation, human experts' supervision would always be required to double check the decision making of these models highlighting the importance of interpretability. In the light of these challenges, this study proposes an interpretable self-supervised multi-task learning model to jointly and effectively tackle the tasks of information retrieval (IR) and extraction (IE) during the current emergency health crisis situation. Our results show that our model effectively leverage the multi-task and self-supervised learning to improve generalization, data efficiency and robustness to the ongoing dataset shift problem. Our model outperforms baselines in IE and IR tasks, respectively by micro-f score of 0.08 (LCA-F score of 0.05), and MAP of 0.05 on average. In IE the zero- and few-shot learning performances are on average 0.32 and 0.19 micro-f score higher than those of the baselines.
\end{abstract}

%% file: 1_introduction.tex
\section{Introduction and Background}
\label{sec:introduction}

The pandemic has resulted in a rapidly evolving literature of scientific publications regarding the novel Coronavirus which has caused an information crisis \cite{roberts2020trec}. Therefore, researchers, healthcare practitioners, policy makers, and other individuals fighting against COVID-19 require specialized information retrieval and extraction systems to keep up with the literature \cite{esteva2020co,wang2020covid}. In this regard, many models are proposed based on state-of-the-art natural language processing (NLP) techniques, including neural ranking models for document retrieval \cite{zhang2020covidex,esteva2020co,koksal2020vapur}, and pre-trained transformers for automated annotation and information extraction \cite{colic2020annotating,lymperopoulos2020concept,spangher2020enabling}.

However, the limited amount of labeled data and drastic changes of COVID-19 landscape \cite{shokraneh2020lessons} has made it challenging for such models to effectively scale up to the domain-specific environment of the current pandemic. Moreover, due to these challenges, such informatic tools cannot manage the pandemic situation on their own, and human experts' supervision is required to double check their decision making. Therefore, it is significantly important for these models to provide textual evidence from the raw input such that human experts can interpret causes of a certain decision making \cite{jin2018attentionmesh,xun2019meshprobenet}. 

Although these challenges are inevitable in every health crisis situation and may be confronted again in near future, very few studies have mainly focused on addressing these specific challenges of the pandemic. In the light of this, we propose a specilized information retrieval and extraction approach which provides interpretability, data efficiency and robustness to the dataset shift problem thereby suitable for the context of the pandemic. Our approach is based on transformers encoders with global-local attention mechanism \cite{beltagy2020longformer}, thus providing interpretability both at document level and sub-word level. We train the encoder in a multi-task fashion alongside a self-supervised learning task which have shown promising results in improving data efficiency, generalization and robustness to dataset shift problems \cite{hendrycks2019using,sun2019test}.



Inspired by \cite{raffel2019exploring} and to exploit the landscape of transfer learning in an optimum way, we design a unified retriever-ranker framework to simultaneously and effectively train our model on a masked language modeling (MLM) self-supervised learning (SSL) task \footnote{We introduce a self-supervised task of masked language modeling which is consistent with our retriever-ranker general framework and enables the model to acquire knowledge about the current state of the pandemic.} along with two supervised tasks: i) information retrieval: retrieval of related articles to a given question, and ii) information extraction: extraction of semantic indexes of PubMed articles which are manually indexed by human experts. Using an input/output transformation module, we translate the tasks of masked language modeling and semantic indexing to the domain of retriever-ranker. For every task, our model gets an input of a given query, a COVID-19 article or question, and initially, retrieves a set of candidate related articles. Next, a transformer-based ranker assign scores to the candidate articles using the attention mechanism proposed by \cite{beltagy2020longformer}, both within and across articles. The final output is computed by transformation of these scores to their corresponding domain. The attention mechanisms within/across the input query and candidate articles enables the model to associate textual evidence from the input to the output decision, thus providing interpretability both on document and sub-word level.

\begin{itemize}
    \item We propose an interpretable, self-supervised, multi-task learning methodology to effectively tackle the tasks of information retrieval and extraction in the context of pandemic. 
    
    \item Our study sheds light on the importance of interpretability and transfer learning--specifically multi-task and self-supervised learning--in addressing the intrinsic challenges of a health crisis situation like the ongoing pandemic; i.e. dataset/covariate shift, lack of unlabeled data and uncertainty. 

    \item We devise a novel mechanism to simultaneously train a unified retriever-ranker on a self-supervised task \textit{masked language modeling (MLM)} and an information extraction (IE) task \textit{semantic indexing} along with an information retrieval (IR) task of retrieving relevant articles to a given question. This enables inter-document representation learning and interpretability which we show is necessary for the context of the pandemic.
\end{itemize}

%% file: 3_methodology.tex
\section{Methodology}
\label{sec:methodology}
In our methodology, we attempt to address the three tasks of self-supervised learning (henceforward we refer to as SSL, or \textit{task\_0}), information extraction (henceforward we refer to as IE, or \textit{task\_1}) and information retrieval (henceforward we refer to as IR, or \textit{task\_2}) using a unified retreival-ranker framework. Our unified framework includes the following three steps which are identical for each task: i) an initial retrieval step which works as a weak classifier\cite{jin2018attentionmesh} to retrieve candidate articles, ii) an I/O transformation technique to map different formats of data into a unified I/O format suitable for next step, and iii) a ranker which works based on transformer encodings of the candidate retrieved articles.

In the context of our retriever-ranker framework, the three aforementioned tasks are defined as follows:

\textbf{Task\_0: (SSL)} A self-supervised pre-text task similar to Masked Language Modeling in \cite{devlin-etal-2019-bert} is performed to introduce knowledge about the context of pandemic. However, in this task, the masked article is treated as a query and masked tokens are selected from a list of covid-19 related terms\footnote{We have used the list of related terms published by NLM \url{https://www.nlm.nih.gov/pubs/techbull/nd20/nd20_mesh_covid_terms.html}}. The model attempts to detect similar articles which include the masked term/s rather than the term itself (order is not important). As such, not only the pre-text task is consistent with the downstream tasks, i.e. the general architecture of our multi-task learning model, but also it effectively learns context matching using both intra- and inter-document information \cite{cohan2020specter}.

\textbf{Task\_1: (IE)} Extraction of semantic indexes assigned to a given article \footnote{Semantic indexing is the information extraction task we address in this study. We use the terms "semantic extraction," "extraction of semantic indexes" and "semantic indexing" interchangeably. In this project, we deal with PubMed articles whose indexes are called Medical Subject Headings (MeSH)}. The given article is treated as a query and outputs are the semantic indexes.

\textbf{Task\_2: (IR)} For a given question as a query a list of related articles are retrieved which include the target answer.

\subsection{Initial Retrieval}
\label{sec::initial-retrieval}

To reduce the number of negative examples, we use an initial retrieval system to retrieve a subset of related articles along with their task specific annotations (for example for extraction of semantic indexes, those that have manually been annotated are considered, and their annotation as candidates for the given task). Our initial retrieval system includes a  document-level embedding model of SPECTER \cite{cohan2020specter}, and Bag-of-Words representation fused with TF-IDF/BM25 following the schema of \cite{jin2018attentionmesh} and \cite{esteva2020co}. SPECTER is initialized with SciBERT \cite{beltagy2019scibert} and trained on a bipartite graph of citations to capture document-level relatedness and minimize a triplet loss between related articles and maximize over unrelated ones. We pre-train SPECTER on PubMed articles and fine-tune it on COVID-19 dataset exclusively. 

In addition, we use TF-IDF/BM25 weighted sum of article tokens to compute a keyword-based representation as well. As such every article is represented as follows:


\begin{equation}
    d = \frac{\sum_{i=1}^{n}{\mathrm{TF-IDF/BM25}}({w}_i,d) \times v_{w_i}}{\sum_{i=1}^{n}{\mathrm{TF-IDF/BM25}}({w}_i,d)} 
\end{equation}

where, $w_i$ is the $i^{th}$ word in article $d$, and $v_{w_i}$ is the token embeddings from the pre-trained model.

We concatenate both representations for every article and query in our database. Next using cosine similarity scores between the input query and other ones, we find the $K$ relevant articles. As for the extraction of semantic indexes (IE task), and masked language modeling, we further retrieve the candidate indexes and mask terms respectively. In this regard, we use the summation of their IDF weights as the scoring scheme to rank and retrieve the candidate indexes and terms. The top $M$ are considered as candidates and passed to the next stage.

\subsection{I/O Transformation}
\label{sec::io-transformation}
In our multi-task learning methodology, the data for every task is fed to a transformer-based ranking system, i.e. our ranker, which is identical for different tasks. Therefore, we require an I/O transformation module to map every task's data format to the unique format of the ranker. The input and output to our ranker is as follows:

\textbf{Input:} given question or article as a query and a list of candidate articles

\textbf{Output:} likelihood scores corresponding to every candidate article

As for the IR task, the input and outputs are consistent with our ranking algorithm. The input is a question along with a list of candidate related articles, and the output for every candidate article is the likelihood score of including the target answer.

In our self-supervised task, the input is the masked articles and a list of candidate articles which include the masked terms withing their contents. Similarly for the information extraction task, the input is a given article and a list of candidate articles which are manually annotated in PubMed.

In our SSL task, the goal is to predict the masked tokens based on likelihood scores of the candidate terms. In our IE task, also, the goal is to extract the semantic indexes of a given article based on likelihood scores of the candidate indexes. However, the output of our ranker assign likelihood scores to the candidate articles instead. Therefore, to train our model for SSL and IE tasks, we need to transform the likelihood scores of candidate terms and indexes, respectively, to those which correspond to candidate documents:
\begin{equation}
\label{eq:transformation}
    Y_{D_c} = T \times Y_{M_c},\hspace{2mm} T \in \mathrm{R}^{K \times M}
\end{equation}
where $Y_{M_c}$ are the likelihood scores of the candidate terms or indexes $M_c$. $T$ is the transformation matrix whose every row is bag-of-word representation of the corresponding document based on the candidate terms. $Y_{D_c}$ are the scores assigned to candidate documents. While choosing candidate documents, we make sure the matrix $T$ is reversible so that the likelihood scores of the terms can be extracted and fed to the loss function. In the task of extraction of semantic indexes task, during inference, the inverse of transformation matrix $T^{-1}$ is used to compute the indexes' likelihood scores which are then passed through thresholds to output the final indexes. Thresholds are optimized to maximize micro-f1 score \cite{pillai2013threshold}.

\subsection{Transformer-based Ranking}
\label{sec::transformer-based-ranking}

Similar to other studies that apply BERT to biomedical \cite{lee2020biobert} or scientific articles \cite{beltagy2019scibert}, query and the candidate articles are tokenized into frequent sub-words using WordPiece unsupervised algorithm \cite{wu2016google} and the original vocabulary of BERT. The sub-word tokenization can also mitigate the problem of out-of-vocabulary words which happens more often during the pandemic. 

Query and the paired articles are separated by [SEP] token and [CLS] token is added to the beginning of every article which is leveraged as the vector representing the corresponding article specified by the query and other relative articles. The same embedding scheme of BERT is used to compute the representation of every input token--summing the token, segment\footnote{an additional embedding is used to differentiate tokens of the query with those of the candidate articles} and position embedding.


Every candidate document ${D_i}^{n-1}$ is encoded along with the paired query $Q^{n-1}$ through multiple layers of a pre-trained transformer \cite{vaswani2017attention}. 

Every candidate article's tokens are encoded, ${D_i}^n$, through a self-attention between its own tokens and a cross-attention between them and the tokens of the query. The [CLS] token of other articles also attend the [CLS] token to enable capturing the inter-document information. Since the length of the input sequence is high in the setting of our methodology, we use global-local attention mechanism \cite{ravula2020etc} to mitigate the computational limitation of transformers in dealing with long sequences. As for the article tokens, we use windowed dilated attention mechanism proposed by \cite{beltagy2020longformer} as a local attention. The attention between [CLS] tokens are also windowed dilated since it is not necessary that all candidate articles attend to every article in order to capture the inter-document relationship. As for the query tokens, however, we use global attention mechanism introduced by \cite{gupta2020gmat}--in which a global memory is utilized that enables every token in the query to attend to every token of candidate articles--to bring in task-specific flexibility.

The query also gets encoded, $Q^n$, using a full self-attention mechanism within its own tokens and a cross-attention between them and the [CLS] tokens of the candidate articles.




%% file: 4_experiments.tex
\section{Experiment}
\label{sec:experiments}
\input{1_table.tex}
\input{1_configuration.tex} 
\subsection{Dataset}

Prior to training on COVID-19 datasets, the models are initially train on general IE and IR datasets. For IE, we use BioASQ's \cite{283} Task 8a dataset which includes almost 15 million article abstracts and titles manually annotated in PubMed. We select 8M recent articles published from 2007-2019. For IR, we use BioASQ's Task 8b dataset which includes 3,243 question paired with related article abstracts. Test sets of each task is used for hyperparameter tuning. We also make sure there is no overlap between these general sets and their corresponding COVID-19 ones.

After the initial training, The models are trained and evaluated on the following three COVID-19 datasets corresponding to the three tasks in hand. As for our self-supervised task, we use CORD-19 dataset \cite{Wang2020CORD19TC} which includes 200K research articles about Coronavirus published in peer-reviewed venues and archival services such as bioRxiv\footnote{\url{https://www.biorxiv.org}} and medRxiv\footnote{\url{https://www.medrxiv.org}}. 

We select CORD-19 articles whose MeSH indexes are manually annotated in PubMed for our semantic indexing task. We crawl and append their MeSH indexes. Consequently, our semantic indexing dataset contains 17K articles which we chronologically sort and split to 13.6K for training  (the oldest 80\%) and 3.4K for testing (the latest 20\%). 

Note: for self-supervised and semantic indexing, we only use the title and abstracts of the articles.

As for IR in the context of COVID-19, we choose TREC-COVID\cite{roberts2020trec} dataset which is an information retrieval dataset for question answering similar to BioASQ QA task phase b. It includes 50 topics, as queries, represented by tuples of (concept, question, narrative). In also includes a dataset of 191K candidate documents from CORD-19. The relevance of 69,317 topic-document pairs are manually evaluated by experts and annotated with three labels of unrelated, partially related, related.



\input{2_table.tex}

\subsection{Evaluation Metrics}

To evaluate the performance of IE we use two sets of evaluation measures: i) flat measures such as micro- and macro-f1 scores, and ii) hierarchical: for which we leverage BioASQ suggested algorithm\footnote{\url{https://github.com/BioASQ/Evaluation-Measures/tree/master/hierarchical}} Lowest Common Ancestor F-measure (LCA-F) \cite{kosmopoulos2015evaluation}.

For evaluation of IR we leverage trec\_eval, the evaluation metrics and algorithms provided by TREC-COVID \footnote{\url{https://trec.nist.gov/trec_eval/}}. The evaluation metrics includes normalized discounted cumulative gain (nDCG@N), P@N, Mean Average Precision (MAP), and Binary preference (Bpref) (see \cite{esteva2020co} for more details on these metrics).

\subsection{Analysis}

In this section, we analyze various elements of our proposed methodology architecture which is essentially retriever + transformer-based ranker (henceforward we refer to as \textit{R+TR}). In this regard, we only use the general datasets, and do not use any sample from COVID-19 datasets for our design decisions or hyperparametes tuning. Table \ref{table:configuration} shows the hyperparameters we have tried and selected for our methodology.

We start with the retriever step which has two hyperparameters for IE task; the number of retrieved candidates $K$, and $M$ the number of candidate MeSH terms; and one for IR task; the number of retrieved candidates $K^{'}$. We use different values and choose those that yields the highest recalls. Since this step is not trainable we can choose different values for $K$ and $K^{'}$. It would only change the size of input and output, not their formats. As shown in Table \ref{table:configuration}, for IE we choose $K=M=512$ because not only it gives the highest recall but also we need $K=M$ so that the $T$ matrix in Equation \ref{eq:transformation} is reversible and likelihood scores of documents can be mapped to those of MeSH indexes during inference. For IR we choose $K^{'}=1024$.

\textbf{Ablation Study}  Moreover, we analyze several design decision for our transformer-based ranking system, the effect of multi-task learning on the general datasets, and experimentally compare our use of different attention mechanisms. The ablation test results for IE and IR are reported in Table \ref{table:ab-semantic-indexing} and Table \ref{table:ab-question-answering} respectively. 

We use two versions of our ranker, a base version (4 layers, 256 hidden size, 8 heads) and a large version (6 layers, 512 hidden size, 8 heads). We evaluate different attention mechanism on the base model. The implementation of full attention mechanism requires truncating the input documents resulting in poor performance. However, the combination of global and dilated sliding window with increasing window size shows better performance than other combinations in both IE and IR. The multi-task learning improves the performance of IR without affecting the IE's performance. Such improvement is expected not only because of the effect of transfer learning but also because IE task is design to improve retrieval and reinforce the latent space to be closer to those of the semantic indexes which human experts believed to be a better representation. Such improvement is relatively higher for larger version of our ranker, showing that the knowledge transfer capability increases with the size of the transformer model. Note: the proposed self-supervised task is exclusively for COVID-19 datasets and disregarded in our ablation analysis. However, its effect is analysed in the following sections.

\input{3_table.tex} 

\subsection{Information Extraction on COVID-19}

In this section, we discuss about the information extraction performance of our methodology exclusively on COVID-19. Table \ref{table:evaluation-semantic-indexing} shows the IE performance of our models and baselines on COVID-19 IE testing set. Results on the left side shows the performance of the models once trained on COVID-19 IE training set. The baselines are similarly fine-tuned with the training data. Our proposed R+TR perform similar to the state-of-the-art baselines without but the inferior without leveraging the proposed self-supervised task and multi-task learning with IR. However, each of these transfer learning techniques significantly improves the IE performance. Leveraging the self-supervised learning task contributes more since R+TR models get acquainted with the context of novel pandemic and it's distributions. Our large model with both self-supervised and multi-task learning achieves higher scores than the baselines by 0.05 LCA-F score and 0.08 Micro F1 score.

The right side shows the performances based on the size of the training data. We chronologically sort the data and train the IE models with the a proportion of the from the beginning. As shown in Table \ref{table:evaluation-semantic-indexing} the partitions include:  0\% which represents the zero shot learning ability, 5\%, 10\%, and 20\% denoting the few-shot learning. Our large R+TR's zero-shor micro-f1 score is significantly higher than the baselines, by 0.32 on average. It achieves 97\% of its optimum performance by using only 20\% of the training data.

\subsection{Information Retrieval on COVID-19}
Table \ref{table:evaluation-qestion-answering} shows the information retrieval performance of our models evaluated on TREC-COVID round 5 dataset. The result for other benchmarks are announced scores for round 5 \footnote{\url{https://ir.nist.gov/covidSubmit/archive.html}}. Our models which are only trained on BioASQ QA task shows inferior performance which is due to the inconsistencies between two tasks, e.g. different distributions of related/unrelated documents for every query. However, leveraging SSL and multi-task learning, our base model beats the top nDCG@20 and Bpref scores. This shows how the proposed transfer learning framework improves model's ability to scale up to a new domain. Our large R+RT achieves significantly superior performance in every metric score.

Furthermore, to analyze the zero- and few-shot learning ability of our model, we fine-tune our SSL multi-task learning models with TREC-COVID dataset. In this regard, we choose round 3 dataset for training which has 40 topics identical to the first 40 topics in round 5. This is because the competition stated from 30 topics in round 1 and every time added 5 topics for the next round. We leave the last 10 topics of round 5 for evaluation. 

We also expand the 40 samples of \textit{(topic , set of candidate documents)} to 1,530 samples by randomly selecting a subset of 128 candidate documents for every given topic rather than 1024. As shown in the bottom of Table \ref{table:evaluation-qestion-answering}, our base and large model are able to effectively leverage such fine-tuning and achieve significantly better scores than the top ones, by 0.05 MAP score. Note that in TREC-COVID challenge also participants could use results from previous rounds.

\input{1_attention-weights.tex}

\subsection{Interpretability}

As mentioned in Section \ref{sec:introduction}, the COVID-19 infodemic has entangled automatic information retrieval and extraction models, so much so that human expers' supervision would always be required. The interpretability can help human experts comprehend the decision making of a model and what has caused a mistaken output. As shown in Figure \ref{fig:weights-fp}, the local-global attention of our model can assist human experts even when it makes an error by providing evidence for the mistaken output and suggesting other alternatives. The model extract the semantic index of SARS-CoV-2 while the manual annotator believes the article is about the general SARS viruses rather than a specific variant. The highlights shows the global attention between the related articles and the query article, and the local attention withing the query article. The weights are averaged and set to three scalar values, following \cite{sarker2019interpretable}, to make the visualization simple \cite{lei2017interpretable}. As depicted by Figure \ref{fig:weights-fp}, the extraction of SARS-CoV-2 is because of the highly matched context about COVID-19 (the top related article) and the last sentence. However, the global attention provide another related article along with suggestions for the correct index. Knowing these, one can quickly identify and fix the error.
\input{2_attention_weights_overtime.tex}
The interpretability can also help to understand the performance of the model in mitigating the challenges of COVID-19 infodemic. As a case study, to analyze the performance of our model in handling the shift in the topics and terminologies of COVID-19 related literature, we take a look at attention weights between the various stigmatized and standard terms for the novel Coronavirus over the time. The stigmatized terms include those which have been used prior to the provisional standard term "2019-nCoV", such as "Wuhan Coronavirus," "Chinese Virus," "Wuhan Novel Pneumonia" to name a few \cite{hu2020covid}. We use the aggregated weights\footnote{Summed and averaged over all sample queries and candidate articles, using both local and global attentions.} when these terms attend to or get attended by the standard ones (i.e. COVID-19 and SARS-CoV-2). We use the chronologically sorted dataset and looked at the weights as the model gets trained over the different time frames.

As shown in Figure \ref{fig:weights-dist}, as the distribution of terminologies changes over time, the attention mechanism learns to relate to the well-established terms mitigating the effect of dataset shift. In the beginning, the model shows high attention weights toward SARS-CoV as it is another variant of Coronavirus which has also originated from Chine; thereby substantially matches with the new context. Then it quickly starts to related stigmatized terms even prior to the introduction of their standard terms. With the advent of the standard terms, the model pays less attention to stigmatized and provisional terms. The attention over SARS-CoV-1 and other related variants decreases as the model dissolves the confusion between them.

%% file: 1_table.tex
\begin{table*}[h!]
\small
\begin{subtable}{0.44\linewidth}
\small
    \begin{tabular}{l | c} \hline
        \centering
        {\textbf{Model}}                                         & {\textbf{MiF}} \\ \hline
        {Medical Text Indexer (MTI)} (default)          & 0.6578  \\
        {MTI} (first line indes)                        &  0.6491  \\ 
        {average top score}                             & \textbf{0.7143}\\ \hline
        {R+TR} (full attention)                         &  0.5532  \\ 
        {R+TR} (increasing $w$) (from 32-512)           &  0.6277 \\ 
        {R+TR} (fixed $w$) (=230)                       &  {0.6142} \\ 
        {R+TR} (decreasing $w$) (from 512-32)           &  {0.6002}\\ 
        {R+TR} (increasing $w$) (dilation on 2 heads)   &  {0.6334}\\ 
        {R+TR} (global + dilated sliding window*)       &  0.6602  \\ \hline
        {R+TR} (base) (w/o multi-task)                  &  0.6602  \\ 
        {R+TR} (base) (w/ multi-task)                   &  0.6671 \\ 
        {R+TR} (large) (w/o multi-task)                 &  0.6980 \\ 
        {R+TR} (large) (w/ multi-task)                  &  \textbf{0.7054}\\ \hline

    \end{tabular}
    \caption{} 
    \label{table:ab-semantic-indexing}
\end{subtable}
\hspace{10mm}
\begin{subtable}{0.45\linewidth}
\small
    \begin{tabular}{l | c} \hline
        \centering
        {\textbf{Model}}                                & {\textbf{MAP}} \\ \hline
        {average top score}                             & \textbf{0.4638}\\ \hline
        {R+TR} (full attention)                         &  0.1911  \\ 
        {R+TR} (increasing $w$) (from 32-512)           &  0.2933 \\ 
        {R+TR} (fixed $w$) (=230)                       &  {0.2805} \\ 
        {R+TR} (decreasing $w$) (from 512-32)           &  {0.2581}\\ 
        {R+TR} (increasing $w$) (dilation on 2 heads)   &  {0.3032}\\ 
        {R+TR} (global + dilated sliding window*)       &  0.3276  \\ \hline
        {R+TR} (base) (w/o multi-task)                  &  0.3276  \\ 
        {R+TR} (base) (w/ multi-task)                   &  0.3441 \\ 
        {R+TR} (large) (w/o multi-task)                 &  0.3550 \\ 
        {R+TR} (large) (w/ multi-task)                  &  \textbf{0.4097}\\ \hline

    \end{tabular}
    \caption{} 
    \label{table:ab-question-answering}
\end{subtable}
\caption{Information extraction (a) and retrieval (b) performances of our models along with the baselines (best performing models of BioASQ Task 8a for IE, and Task 8b Phase A for IR). The baseline scores are the average of their provided Micro-F1 and MAP (respectively for IE and IR) scores across all test batches. We also evaluate our models on all the provided test batches. * dilated sliding window uses increasing window from 32 to 512 and dilation on 2 heads.}
\end{table*}

%% file: 1_configuration.tex
\begin{table}[h!]
\centering
    \begin{tabular}{l c}
    \hline
\textbf{Hyperparameter}   &    \textbf{Value(s)}    \\ \hline
$ |V| $        &    20M, \textbf{30M}    \\
$K$       & 128,  256 , \textbf{512}, 1024    \\
$M$       &  128, 256, 512, \textbf{1024}    \\
$w$      & 32,..., 512, $\mathbf{inc}{[32:512]}$, $\mathrm{dec}{[32:512]}$ \\
dilation      &  0, 1, 2, 3, $\mathbf{inc}{[0:3]}$ \\
dilation heads & 1, \textbf{2}, 3 \\
dorpout   &   0.1, \textbf{0.2}, 0.3, {\textbf{0.4}$^*$} \\ 
batch size   & 8, \textbf{16}, 32, 64 (gpu memory limit) \\
output vector size      & 512, \textbf{1024}, 2048\\
w.e. size   &   128, \textbf{256}, {\textbf{512}$^*$} \\ 
hidden size       &  128, \textbf{256}, {\textbf{512}$^*$}  \\ 
\#layers       &   \textbf{4}, 5, {\textbf{6}$^*$}, 7, 8    \\ 
learning rate   &  0.001, \textbf{0.0005}, 0.00025,  \textbf{0.0001} \\\hline

    \end{tabular}
    \caption{ Hyperparameters value. w.e.: embedding size for initial retrieval step. We use bold text for the optimal ones among all tried values. $^*$ refer to those for large ranker. Best dilation size is achieved by increasing it by 1 from first layer to the last.} 
    \label{table:configuration}
\end{table}

%% file: 2_table.tex
\begin{table*}
    \footnotesize
    \centering
    \begin{tabular}{l l l l l | c c c c} \hline
        \multirow{2}{40pt}{\textbf{Model}}   & \multirow{2}{40pt}{\textbf{LCA-F}}   & \multirow{2}{40pt}{\textbf{MiF}}  & \multirow{2}{40pt}{\textbf{MaF}}   & \multirow{2}{40pt}{\textbf{Accu.}}  & \textbf{MiF}  & \textbf{MiF}  & \textbf{MiF}  & \textbf{MiF} \\
        & & & & & (0\%) & (5\%) & (10\%) & (20\%)  \\  \hline 
        {MTI} (default)                                          & 0.5630 & 0.7299 & 0.5058 & 0.4913 & 0.2216 & 0.3322 & 0.4592 & 0.5637  \\
        {MTI} (first line indes)                                 & 0.5534 & 0.7220 & 0.5011 & 0.5066 & 0.2185 & 0.3092 & 0.4617 & 0.5774 \\ 
        {attention mesh}                                         & \textbf{0.5788} & \textbf{0.7645} & 0.5298 & 0.5576 & 0.2710  & 0.3965 & 0.5041 & 0.6186 \\ \hline
        {R+TR} (base) (w/o ssl \& mt)                   & 0.5396 & 0.7002 & 0.4924 & 0.4855 & 0.3066 & 0.4334 & 0.5038 & 0.5910  \\ 
        {R+TR} (base) (w/ ssl)                                   & 0.5521 & 0.7282 & 0.5060 & 0.5099 & 0.3804 & 0.4855 & 0.6160 & 0.6634  \\ 
        {R+TR} (base) (w/ ssl \& mt)                    & 0.5632 & 0.7555 & 0.5111 & 0.5234 & \textbf{0.4852} & 0.5920 & 0.6762 & \textbf{0.7241}  \\ 
        {R+TR} (large) (w/o ssl \& mt)                  & 0.5616 & 0.7419 & 0.5023 & 0.5225 & 0.3631 & 0.4743 & 0.5594 & 0.5952 \\ 
        {R+TR} (large) (w/ ssl)                                  & \textbf{0.5974} & 0.7767 & 0.5319 & 0.5688 & 0.4902 & 0.6191 & 0.6978 & 0.7330 \\ 
        {R+TR} (large) (w/ ssl \& mt)                   & \textbf{0.6115} & \textbf{0.8102} & \textbf{0.5587} & 0.5864 & \textbf{0.5644} & 0.6759 & 0.7411 & 0.7888 \\ \hline

    \end{tabular}
    \caption{Information extraction performance of our proposed models in comparison with baselines. ssl: self-supervised learing; mt: multi-task learning. The second half of the table shows the Micro F1 score based on the size of the COVID-19 training dataset ranging from 0\% (zero-shot) to 20\% (few-shot).} 
    \label{table:evaluation-semantic-indexing}
\end{table*}

%% file: 3_table.tex
\begin{table*}
    \centering
    \begin{tabular}{l c c c c} \hline
        model                                                    & \textbf{nDCG@20} & \textbf{P@20} & \textbf{Bpref} & \textbf{MAP} \\  \hline 
        top score                                                & \textbf{0.8496} & \textbf{0.8760} & \textbf{0.6378} & \textbf{0.4731}  \\
        ranke\#1 in nCDG@20                                       & \textbf{0.8496} & \textbf{0.8760} & 0.6372 & 0.4718  \\ 
        ranke\#1 in P@20                                          & \textbf{0.8496} & \textbf{0.8760} & 0.6372 & 0.4718 \\ 
        ranke\#1 in Bpref                                         & 0.8490 & 0.8690 & \textbf{0.6378} & \textbf{0.4731} \\
        ranke\#1 in MAP                                           & 0.8490 & 0.8690 & \textbf{0.6378} & \textbf{0.4731} \\ \hline
        {R+TR} (base) (w/o ssl \& mt)                            & 0.7915 & 0.8383 & 0.6021 & 0.4498  \\ 
        {R+TR} (base) (w/ ssl)                                   & 0.8198 & 0.8556 & 0.6259 & 0.4685  \\ 
        {R+TR} (base) (w/ ssl \& mt)                             & \textbf{0.8571} & 0.8707 & \textbf{0.6420}  & 0.4713  \\ 
        {R+TR} (large) (w/o ssl \& mt)                           & 0.8047 & 0.8489 & 0.6204 & 0.4566\\ 
        {R+TR} (large) (w/ ssl)                                  & 0.8302 & 0.8607 & 0.6331 & 0.4752 \\ 
        {R+TR} (large) (w/ ssl \& mt)                            & \textbf{0.893} & \textbf{0.8911} & \textbf{0.6573} & \textbf{0.4911} \\ \hline
        {R+TR} (base) (w/ ssl \& mt) (w/ f.t.)                   & \textbf{0.8986} & \textbf{0.9152} & \textbf{0.6635}  & \textbf{0.5056}  \\ 
        {R+TR} (large) (w/ ssl \& mt) (w/ f.t.)                            & \textbf{0.9237} & \textbf{0.9459} & \textbf{0.6904} & \textbf{0.5230} \\ \hline

    \end{tabular}
    \caption{Information retrieval performance of our model with and without pre-training on self-supervised and semantic extraction tasks. The performance of SSL models are improved which shows the models learn the current time period context of medicine literature. During the IE task training the model's attention gets well trained to extract the relative semantic indexes from articles which is basically the main part of the IR procedure in bio-medicine articles.} 
    \label{table:evaluation-qestion-answering}
\end{table*}

%% file: 1_attention-weights.tex
\begin{figure*}
    \centering
    \small
    \includegraphics[scale=0.40]{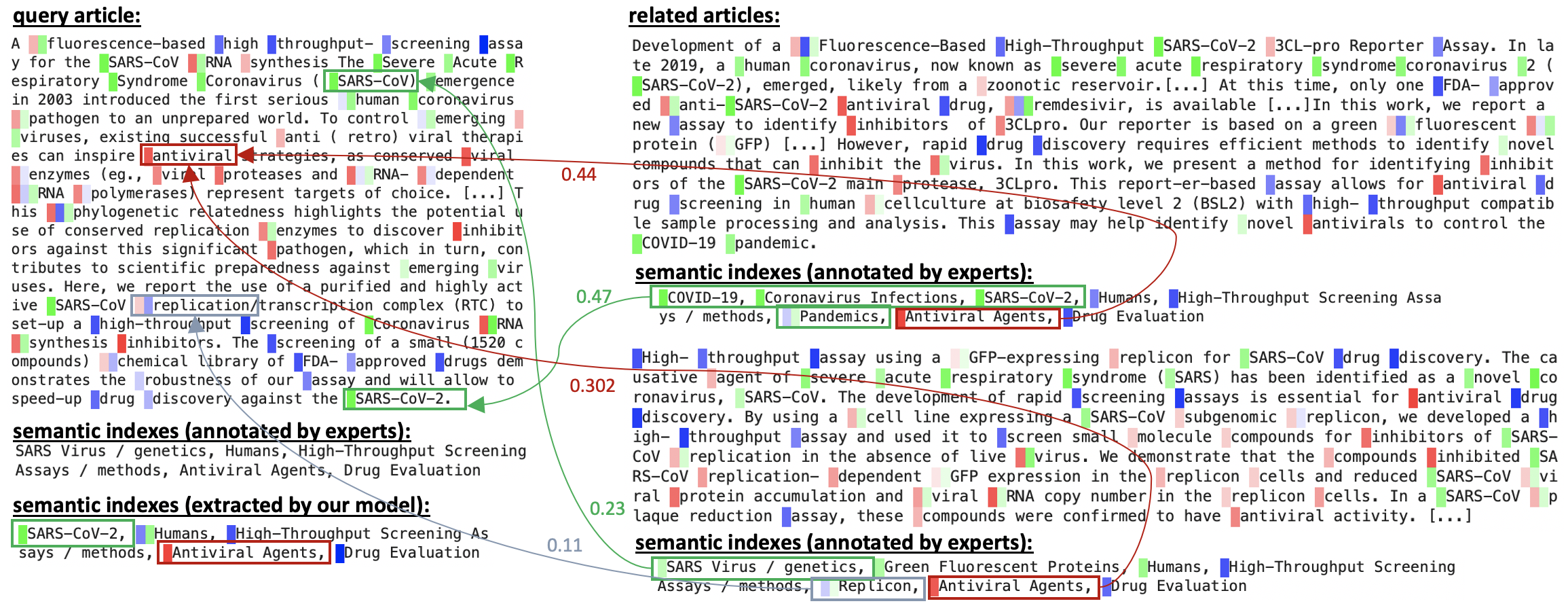}
    \caption{Illustration of attention weights between the input query and candidate articles along with the extracted outputs. The intensity and the color of the highlights denotes attention weights' values which is averaged and set to three scalar between the highly correlated terms.}
    \label{fig:weights-fp}
\end{figure*}

%% file: 2_attention_weights_overtime.tex
\begin{figure}
    \centering
    \small
    \includegraphics[scale=0.44]{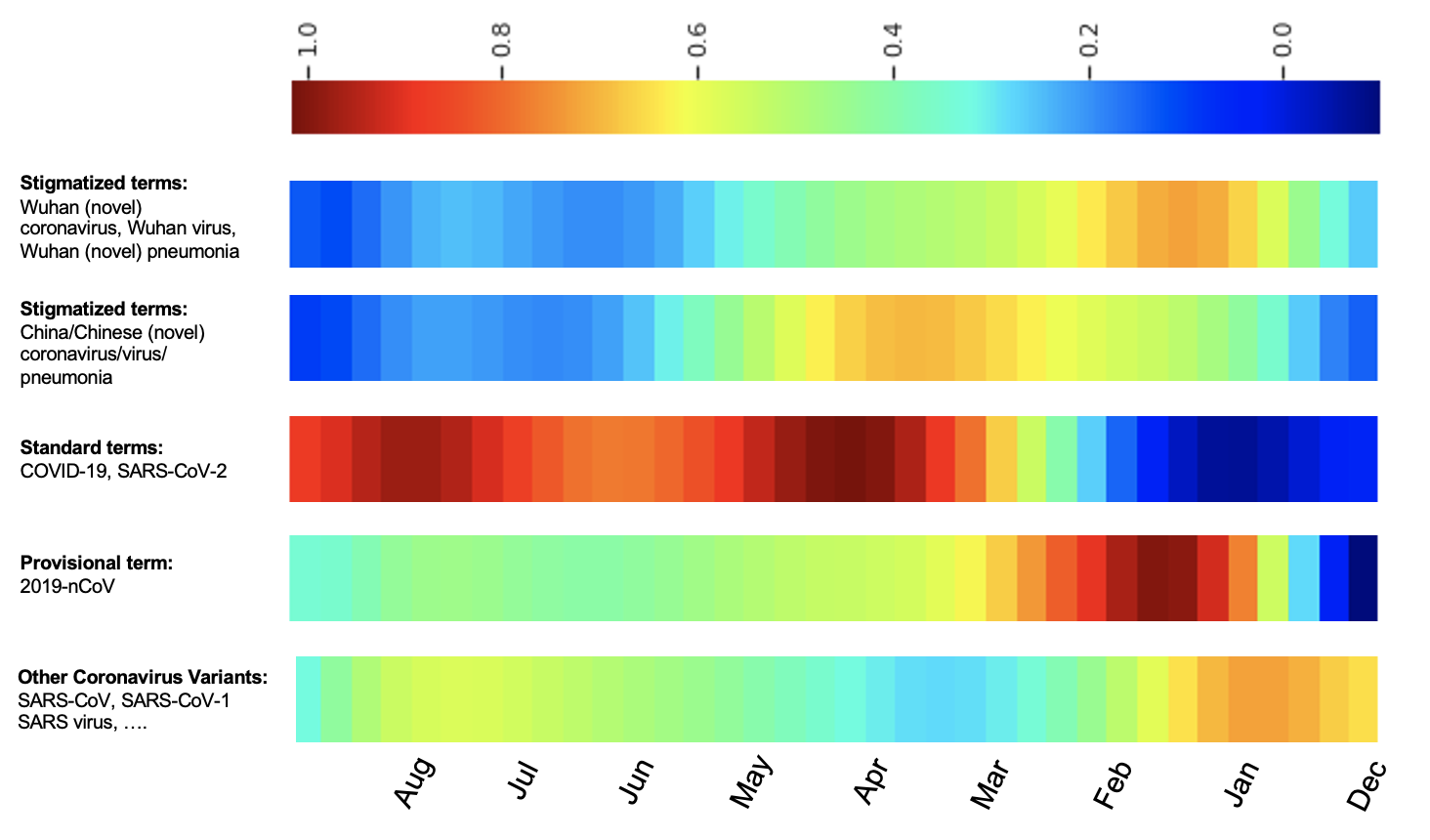}
    \caption{Attention weights of terms attending to COVID-19 and SARS-CoV-2 over different time frames. These weights are normalized for visualization purpose, following \cite{nguyen2019transformers}.}
    \label{fig:weights-dist}
\end{figure}

%% file: 5_conclusion.tex
\section{Conclusion}
In this study, we have unified the tasks of information retrieval for question answering and extraction of semantic indexes along with a self-supervised pre-text task to leverage the advantages of transfer learning in addressing the data efficiency, generalization and dataset shift issues confronting these application in the current pandemic. In comparison to benchmarks, our model learns with fewer number of labeled data and shows substantially higher zero-shot performance. Furthermore, the performance of the tasks which have scarce amount of labeled data improves dramatically once simultaneously trained with other tasks for which relatively decent amount of labeled data is available. 

In addition, the global-local attention mechanism, which is the perfect fit for retrieval and ranking models, provided our model with not only robustness to the drastic changes of the literature, but also interpretibility both on document and token level. We have shown the potentials of such interpretability in understanding the decision making of the model as well as lightening the load on human experts.  

\textbf{Fututre Works:} Our study brings focus towards state-of-the-art remedies to the current challenges of the pandemic which opens up new doors to a more systematic analysis of each of these challenges and more sophisticated algorithms in that regard. In future, we would like to combine more information retrieval and extraction tasks as more data is being annotated and prepared for domain-specific environment of the pandemic. To better evaluate the performance of the global-local interpretability, we plan to perform qualitative analysis by providing this tool to a human expert and analyze their time efficiency and performance. Furthermore, we believe that such unifying mechanism would contribute to more applications and situations other than the current pandemic. We aim to identify other situations which cause drastic changes and growth in the literature and examine how such models would perform in mitigating those challenges.
